\documentclass[iopart,twocolumn]{revtex4}
\usepackage{amsmath,amssymb,bm,graphicx,subfigure,multirow,color}
\begin{document}

\title{Anharmonicity induced resonances for ultracold atoms and their
detection}
\author{J. P. Kestner$^{1,2}$, L.-M. Duan$^1$}
\affiliation{$^1$Department of Physics, University of Michigan, Ann Arbor, MI 48109\\
$^2$ Condensed Matter Theory Center, Department of Physics, University of Maryland, College Park, MD 20742.}

\begin{abstract}
When two atoms interact in the presence of an anharmonic potential, such as
an optical lattice, the center of mass motion cannot be separated from the
relative motion. In addition to generating a confinement-induced resonance
(or shifting the position of an existing Feshbach resonance), the
external potential changes the resonance picture qualitatively by
introducing new resonances where molecular excited center of mass states
cross the scattering threshold. We demonstrate the existence of these
resonances, give their quantitative characterization in an optical
superlattice, and propose an experimental scheme to detect them through
controlled sweeping of the magnetic field.
\end{abstract}

\maketitle
\section{Introduction}
In recent years, there has been much progress in the study of
ultracold atoms in optical lattices, which can cleanly emulate important
models in condensed matter, hold promise for quantum computing schemes, and
offer the prospect to observe many interesting new phenomena \cite{review}.
The versatility of this line of research is due in no small part to the
control of the atomic interactions afforded by tuning an external
magnetic field near a Feshbach resonance \cite{Feshbach}. In addition to a
magnetic field, a confining potential can also be used to tune the
scattering length via a
Feshbach-type mechanism, typically referred to as a confinement-induced
resonance \cite{CIR} or a trap-induced shape resonance \cite{TISR} depending
on the trap configuration. The trap-induced resonance is basically caused by
a shift of the free-space Feshbach resonance point by the
confining potential \cite{Kohl05}. In an optical lattice, the possibility of decay of atomic pairs due to anharmonic coupling to molecules in an excited center-of-mass (c.m.)\ state has previously been discussed in Ref.~\cite{Bolda05}.  More recently, the possibility of a controlled transfer of an atomic pair to a molecule in an excited c.m.\ state was discussed in Ref.~\cite{Mentink09}.  The anharmonicity of the optical lattice potential has also been recognized as important in obtaining quantitatively accurate predictions for the shift of the free space Feshbach resonance position, binding energy, etc.~\cite{anharm2}.

In this paper we point out a new effect whereby anharmonic confinement, e.g., from an optical lattice, not only
shifts the free-space resonance point, but also induces a series of additional scattering resonances.  (A similar effect occurs in mixed dimensions in the absence of anharmonicity \cite{Nishida08}.)  Thus, anharmonicity may give rise not only to population transfer between atom pairs and molecules different c.m.\ states, as mentioned to various extents in previous works \cite{Bolda05,Mentink09,anharm2}, but also to a strong modification of the effective atom-atom interaction in the vicinity of the induced scattering resonances.  In order to use the optical lattice system as a quantum emulator, it is important to have a full understanding of the dependence of the interaction on the experimental parameters.  Even such a basic item as the form of the effective many-body lattice Hamiltonian \cite{lattice} will be affected in the vicinity of an induced resonance (for a detailed treatment, see Ref.~\cite{Kestner09}).  The presence of the additional resonances is then an important consideration for experiment as well as a novel tool for tuning the interaction utilizing a resonance between atoms and excited c.m.\ molecules.  Measuring population transfer between c.m.\ states can provide a handy means to look for the resonances. Below, we characterize these anharmonicity induced resonances in an optical superlattice and propose an experimental scheme to detect their consequences.

\begin{figure}[tbp]
\subfigure{\includegraphics[height=.28\columnwidth,width=1\columnwidth]
        {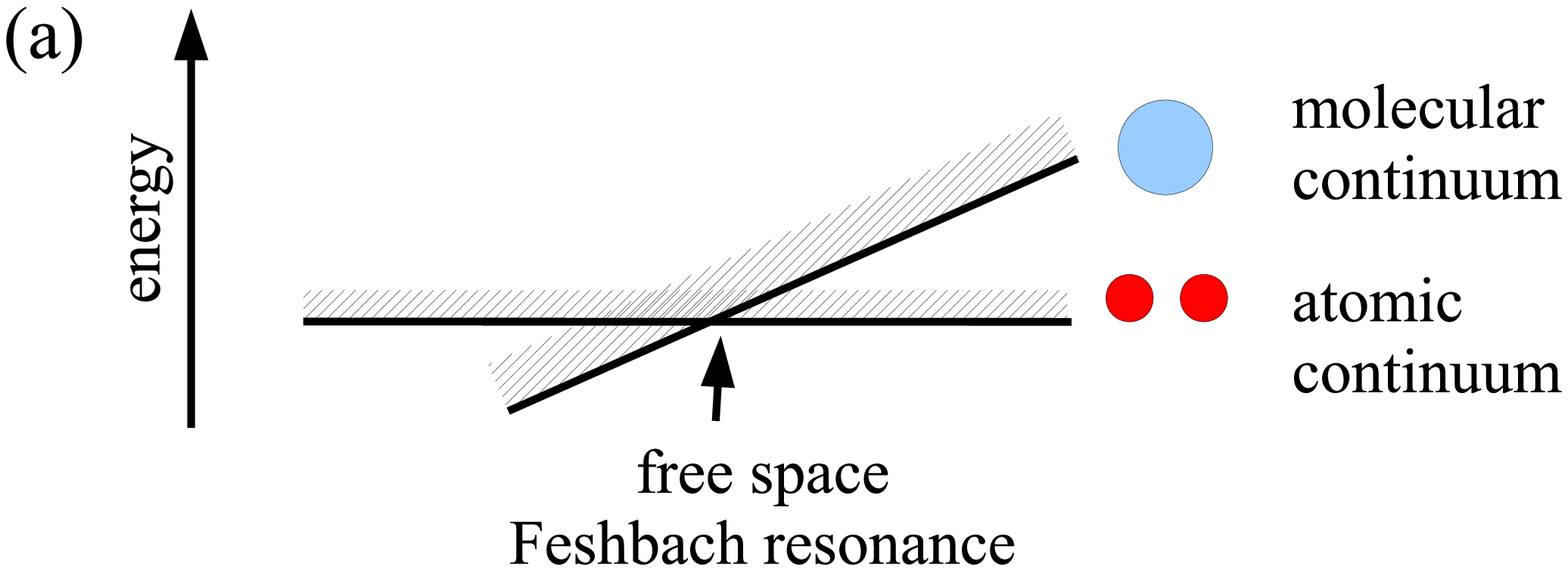}\label{fig:ideafree}}
\subfigure{\includegraphics[height=.34\columnwidth,width=1\columnwidth]
        {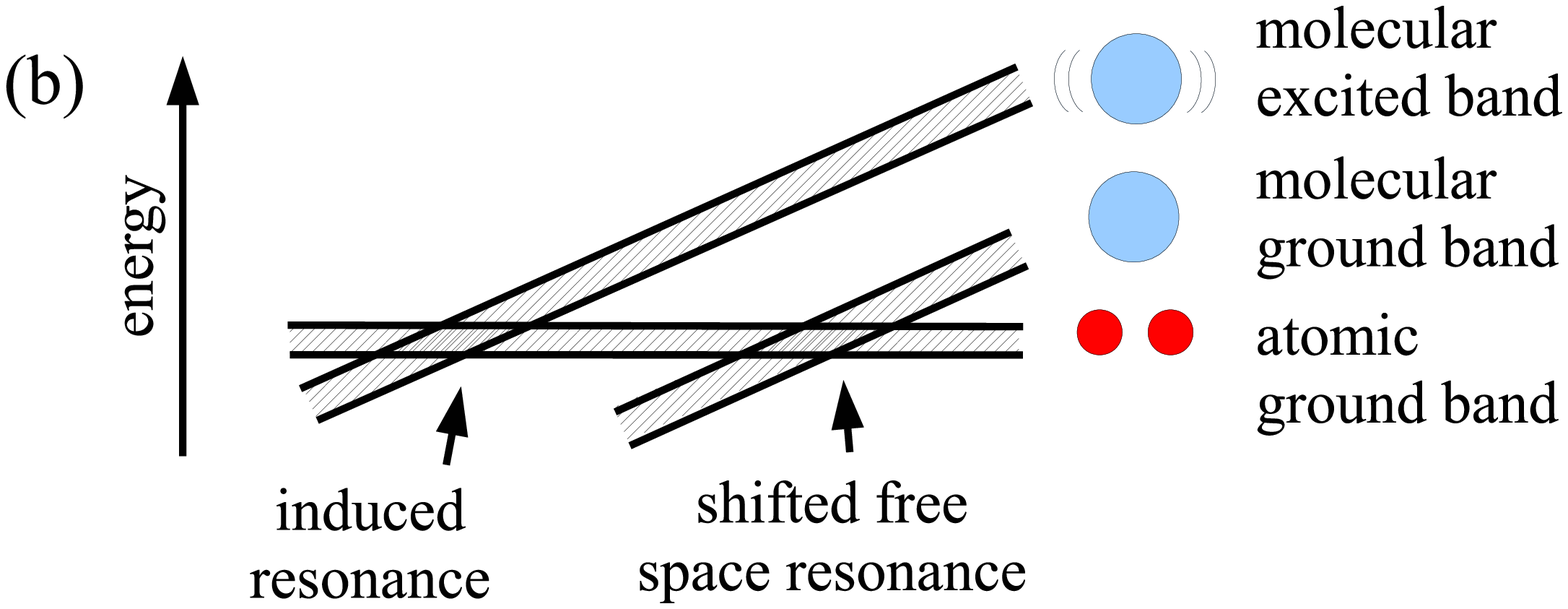}\label{fig:idealattice}}
\subfigure{\includegraphics[height=.4\columnwidth,width=1\columnwidth]
        {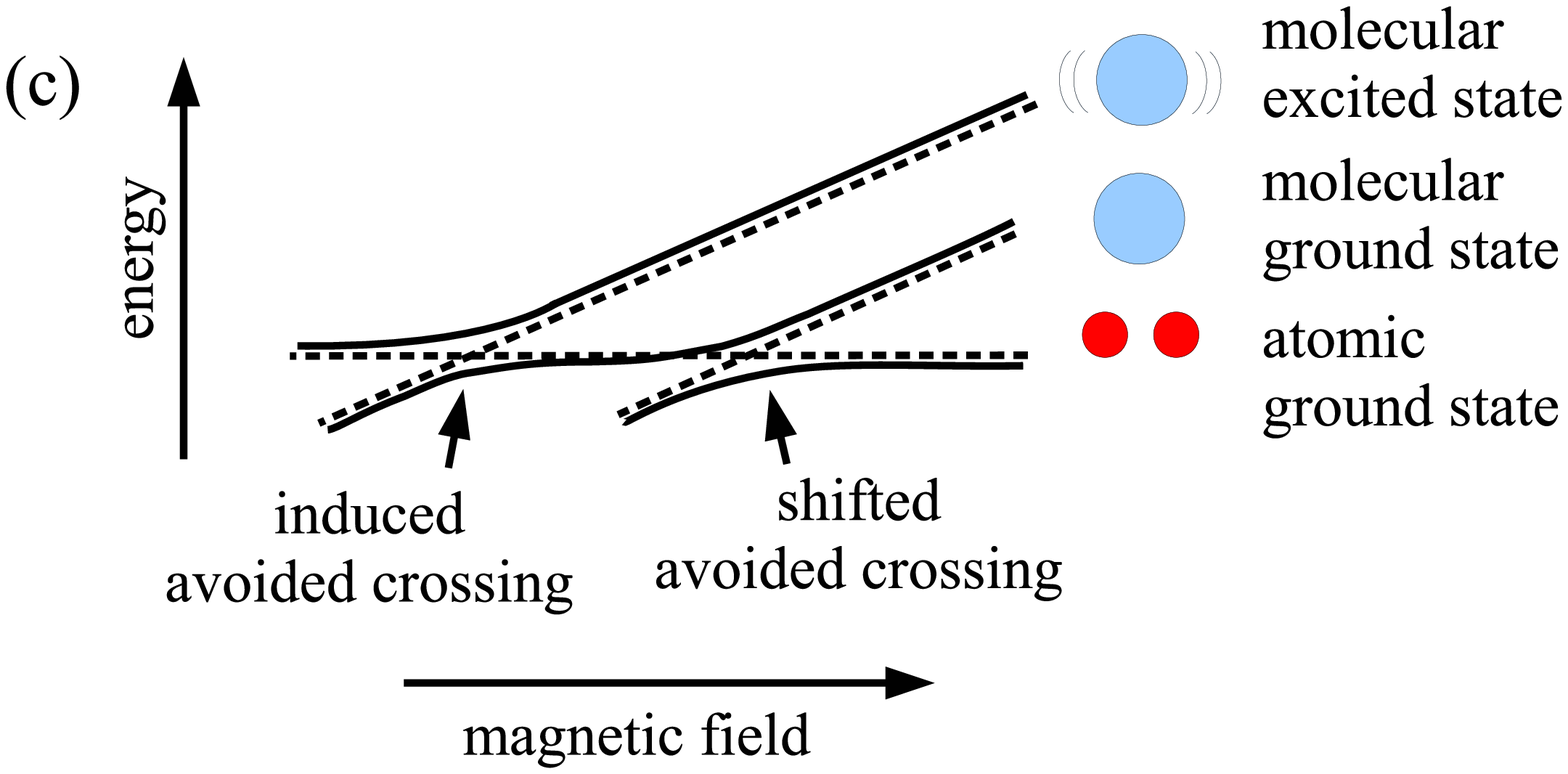}\label{fig:ideatrap}}
\caption{Sketches of the Feshbach type of resonances (a) in free
space; (b) in an optical lattice with additional anharmonicity induced resonances; (c) in a confining potential where the
resonances are signaled by the avoided level crossings.}
\label{fig:idea}
\end{figure}
To understand the basic mechanism of the anharmonicity induced
resonances, let us first compare it with the free-space Feshbach
resonance. The free space Feshbach resonance is caused by coupling
between the scattering state of the atomic pair and a highly
excited molecular level (the Feshbach molecule), as depicted in
Fig.~\ref{fig:ideafree}. When the energy of the Feshbach molecule,
tuned by the external magnetic field, crosses the lowest
scattering state, a resonance in the scattering length is signaled
\cite{Feshbach}. In free space, c.m.\ and relative motions are
decoupled during the atomic scattering, and the
c.m.\ momentum forms a continuum which is not altered by the scattering
process.

In the presence of an optical lattice, the continuum spectrum for the atomic
and the molecular c.m.\ motion both split in a series of energy bands. We
consider scattering of the atoms in the lowest bands, and only this lowest
atomic band is shown in Fig.~\ref{fig:idealattice}. However, even for this
lowest-band atomic scattering, the excited bands for the c.m.\ motion of the
Feshbach molecule still play a significant role due to the
anharmonicity of the optical lattice potential. In a harmonic potential, the
c.m.\ motion of two colliding atoms is separated from their relative motion,
and thus remains in the lowest band during the collision and does not couple
to the Feshbach molecule in the excited bands. However, the anharmonicity of
the potential mixes the c.m.\ and relative motions, and the lowest band
scattering state of the atoms is coupled to the Feshbach molecule in \emph{each band}, as depicted in Fig.~\ref{fig:idealattice}. As one can see from
this figure, all the bands for the Feshbach molecule, no matter how excited,
eventually cross the atomic scattering threshold as the
magnetic field is lowered. This will lead to many resonances for the atomic scattering.
In practice, the anharmonic coupling between a Feshbach molecule in the
excited band and the atomic pair state in the lowest band will decrease as
the band becomes more excited, and the resonances become progressively
narrower as the magnetic field is lowered, so only the first few of these
resonances are broad enough to be experimentally observable.

In order to quantitatively characterize the anharmonicity induced
resonances, we consider the atoms in an optical superlattice potential. In
an optical lattice, direct calculation of the scattering length
between two atoms is challenging as one can not separate the c.m.\ and the
relative motion and solution of an equation with all six degrees of freedom
is numerically demanding. Instead, here we consider the atoms in a deep
superlattice potential \cite{dblwell}, which separates the periodic optical
lattice into a series of double-well potentials. This has several
motivations: First, by adding a confining trap, as illustrated in
Fig.~\ref{fig:ideatrap}, the resonance in the continuum scattering spectrum caused by
the emergence of a new Feshbach molecular level becomes an avoided level
crossing in the discrete spectrum of the trapped atoms. By calculating the
width and the position of the avoided level crossing, we can approximately
characterize the resonance properties for the atomic scattering.
Numerically, it is more convenient to deal with the discrete spectrum in a
trap which allows application of specific calculation techniques presented
below. Second, the optical superlattice potential has been realized in
experiments \cite{dblwell}, which allows direct detection of consequences
of the anharmonicity induced resonances in this kind of trap. We will
propose an experimental scheme to test the quantitative predictions from the
anharmonicity induced resonances in a superlattice. Third, the anharmonicity induced resonances also affect the effective many-body Hamiltonian for
strongly interacting atoms in an optical lattice \cite{Kestner09, lattice}. A natural
step to derive such a Hamiltonian is to first consider the
effective interaction for atoms in double-well potentials realized with
a deep optical superlattice.

\section{Methods}
\begin{figure}[tbp]
\subfigure[]  {\includegraphics[height=.375\columnwidth]
        {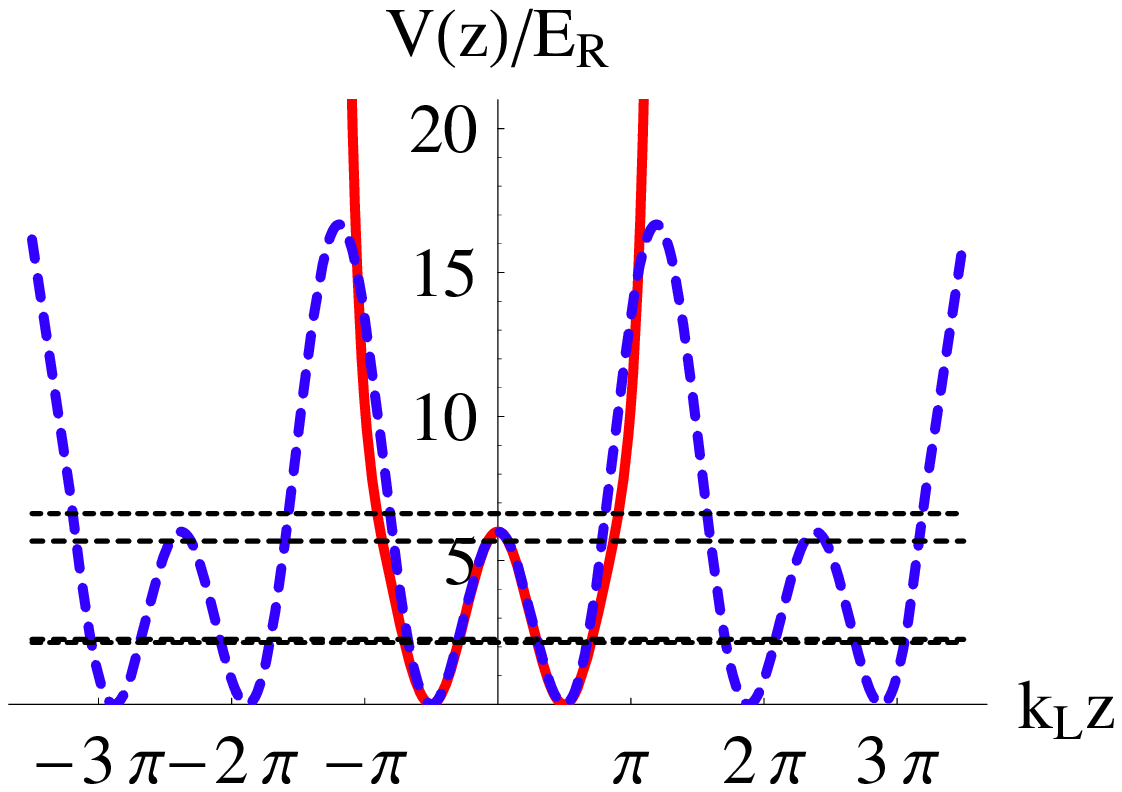}}
\subfigure[]  {\includegraphics[height=.375\columnwidth]
        {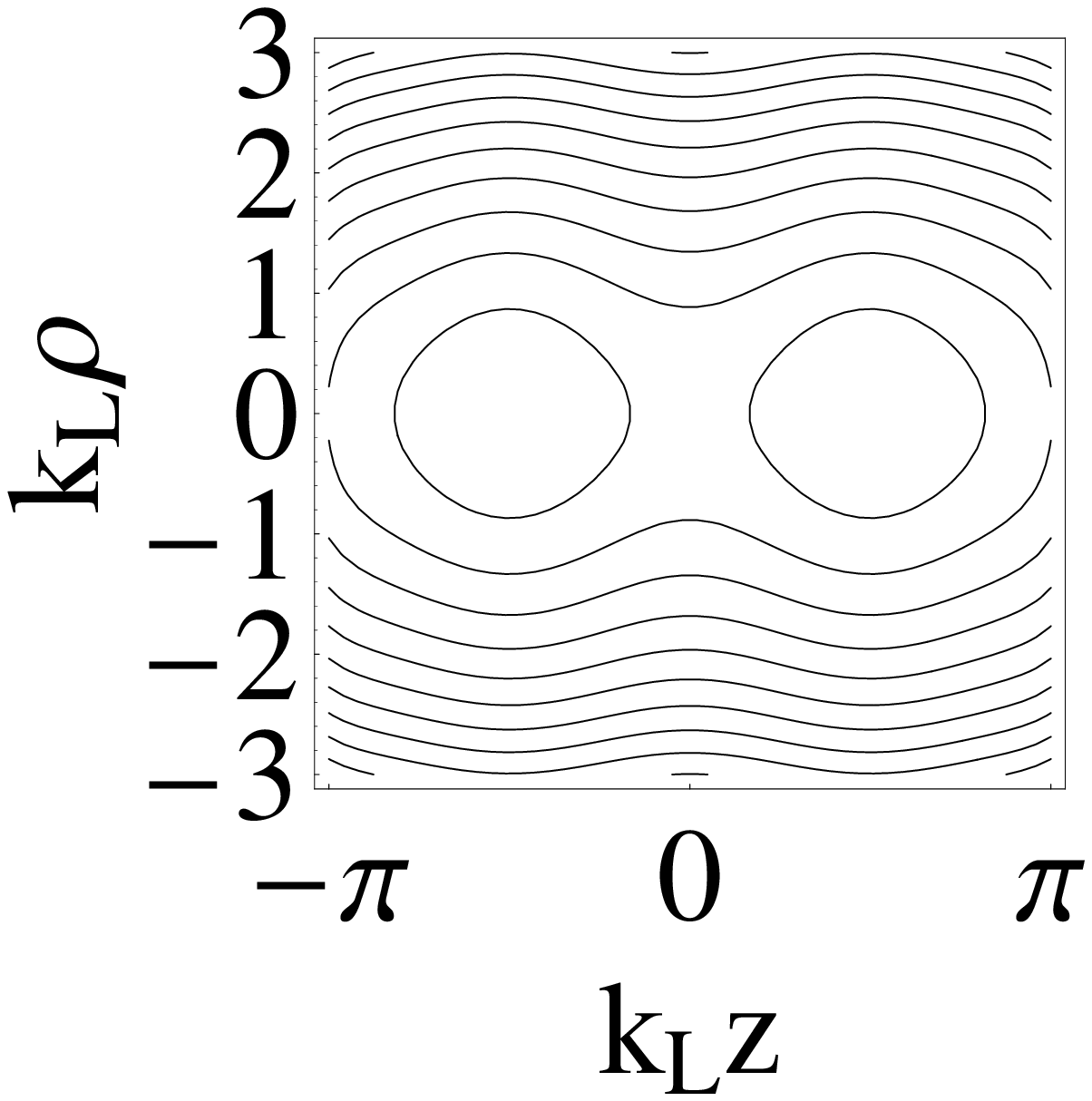}}
\caption{(Color online.) (a) Double-well potential (solid) for a single atom along $z$ and an example of a corresponding superlattice potential (dashed).  The horizontal lines are the lowest four single-atom energy levels of the double-well. (b) Contour plot of the locally isotropic 3D double-well potential.}\label{fig:dblwell}
\end{figure}

We assume that the superlattice potential is along the axial direction $z$
which separates the system into a series of double-wells \cite{dblwell}. We
consider two atoms of mass $m$ in each double-well
potential $V\!(z)$, approximated by Taylor expanding $V_0 \cos ^{2} \left(k_{L}z\right)$
to 12$^{\text{th}}$ order in $z$.  Here $V_0$ sets the barrier depth and $k_L$ sets the distance between wells, and is related to the laser wavevector.  Although it is not important for our purposes to exactly fit a particular form of superlattice potential, if one takes a superlattice of the form $\cos ^{2} \left(k z\right) + c \sin^2 \left(k z/2\right)$, then one should choose $k_L = \frac{\pi k}{4 \arccos \sqrt{1/2+c/8}}$, as shown in Fig.~\ref{fig:dblwell}.  (We express energy in units of a "recoil energy," $E_R = \hbar^2 k_L^2/2m$, and plot the case $V_0 = 6E_R$.)  In any case, this potential should be quite sufficient to capture the essential physics in the limit of independent double-wells. For ease of calculation, the lattice wells in the transverse directions are approximated by harmonic potentials, with the frequency, $\omega $, chosen such that the potential is locally isotropic at the bottom of each well.

Due to the harmonicity of the transverse trap, the c.m.\ motion in the
transverse direction separates out and is thus neglected in the rest of the
discussion. Also, due to the axial symmetry of the trap, the azimuthal angular dependence of the relative motion separates out.  However, along the axis of the double-well, the c.m.\ motion is
not separable from the relative motion. The two atom system then has three
relevant coordinates: the relative coordinates $z$ and $\rho$, along the
axial and transverse directions, respectively, and $Z$, the axial c.m.\
coordinate. In terms of these coordinates, the external potential with
barrier depth $V_0$ is $V\! \left( \rho, z, Z \right) = V\! \left( \rho
\right) + V\! \left( z, Z \right)$, where
\begin{align}
V\! \left( \rho \right) &= V_0 k_L^2 \rho^2/2 \notag
\\
V\! \left( z, Z \right) &= V_0 \sum_{ \substack{ n=0 \\ \pm} }^6 \frac{ \left( -4 \right)^n \Gamma \left( 1 - 2n \right) }{ \Gamma \left( 1- 4n \right) \Gamma \left( 1 + 4n \right) } k_L^{2n} \left( Z \pm \frac{z}{2} \right)^{2n} \notag
\end{align}
with $\Gamma \left(x\right)$ the Euler gamma function, and our summation over signs denotes that for each value of $n$ one must also add the two terms corresponding to the upper and lower signs.

The atoms are interacting via a short-range potential $U\!\left( r\right) $ characterized by its s-wave scattering
length $a_{s}$.  The exact form of the interaction is irrelevant as long as its effective range is much smaller than the average
interatomic distance and the trap length scale.  For a broad s-wave Feshbach resonance, the use of a zero-range pseudopotential is typically justified \cite{Chen07}, as has also been confirmed experimentally \cite{Stoferle06}. Numerically, it is easier to
use a finite-range attractive Gaussian interaction $U\!\left( r\right)
=-U_{0}\exp \left( -r^{2}/r_{0}^{2}\right) $, where we typically take
$r_{0}=0.05\sqrt{\hbar /m\omega }$. Finite-range effects should be negligible for such small values of $r_{0}$, and we
have verified this by repeating our calculations with $r_{0}=0.1\sqrt{\hbar /m\omega }$.  The free space scattering
length is varied by adjusting the strength of the interaction, $U_{0}$.  We have used values of $U_0$ such that the potential supports either zero bound states (for negative scattering length) or one bound state (for positive scattering length).  The scattering length goes through resonance when the lowest eigenstate of the interaction potential passes from being unbound to bound.

Adopting units such that $k_{L}=1$ and $E_{R}=\hbar ^{2}k_{L}^{2}/2m=1$, the
Hamiltonian may be written as
\begin{multline}\label{eq:Hr}
H=-\frac{2}{\rho} \frac{\partial}{\partial \rho} \rho \frac{\partial}{\partial \rho} -2\frac{\partial^2}{\partial z^2} - \frac{1}{2}\frac{\partial}{\partial Z^2}
\\
+2\frac{m^2_{\ell}}{\rho^2}+ V\left( \rho ,z,Z\right) -U_{0}e^{-\left( z^{2}+\rho ^{2}\right)
/r_{0}^{2}}
\end{multline}
where $m_{\ell}$ is the relative angular momentum, which is a good quantum number due to axial symmetry.  In the following we will only consider $m_{\ell}=0$, since, in the limit as $r_0$ goes to zero, the interaction does not affect states with $m_{\ell} \neq 0$.

We find the low-lying states of the system using a stochastic variational
method \cite{svm}. In this approach, the variational wavefunction
takes the form
\begin{equation}
\Psi\!\left( \rho ,z,Z\right) =\sum_{i}^{N}\alpha _{i}\exp\! \left( -\rho
^{2}/a_{i}^{2}-z^{2}/b_{i}^{2}-Z^{2}/c_{i}^{2}\right) ,  \label{eq:psi}
\end{equation}
where $\alpha$ is a linear variational parameter, $\{ a, b, c \}$ are nonlinear variational parameters which define the
basis elements, and $N$ is the size of the basis set. The nonlinear parameters are selected from stochastically
generated pools of candidates to minimize the variational energy $\langle \Psi |H| \Psi \rangle / \langle
\Psi | \Psi \rangle$. The basic algorithm is as follows: starting with a set of $N-1$ basis states,

\begin{description}
\item[1)] a pool of (in our calculations) $25$ new basis states is randomly generated, each defined by a given value
    of $\{a_{i},b_{i},c_{i}\}$;

\item[2)] for each of the $25$ possible $N$-dimensional basis sets formed by adding one basis from the candidate
    pool, the energy is minimized with respect to $\alpha $;

\item[3)] the new basis set that yields the lowest energy is kept and the previous steps are repeated until the
    basis size, $N$, increases to the desired number.
\end{description}

Once every few iterations, the existing basis set is optimized by the following refining process: starting with a set of $N$ basis states and $n=1$,

\begin{description}
\item[A)] a pool of $25$ replacement basis states is randomly generated, each defined by a given value of
    $\{a_{n},b_{n},c_{n}\}$;

\item[B)] for each of the $25$ possible $N$-dimensional basis sets formed by replacing the $n^{\text{th}}$ old basis
    state with a new one from the candidate pool, the energy is minimized with respect to $\alpha $;

\item[C)] if the lowest of these $25$ energies is lower than the current variational energy, the $n^{\text{th}}$ old
    basis state is replaced by the new optimal one and the previous steps are repeated for $n=1...N$.
\end{description}

We typically achieve fairly good convergence for $N\sim 300$. Although in principle the nonlinear basis optimization
must be performed for each value of $a_{s}$, actually the basis set does not change too much as one sweeps across
resonance except to include narrower and narrower Gaussians for positive $a_{s}$ where deeply bound molecules form.
Apart from deeply bound states, the change in the wavefunction is mainly due to changing the expansion coefficients,
$\alpha $. To save computational time then, we performed the nonlinear basis optimization for four different values of
$ a_{s}$ ranging from positive to negative, joined the four optimized basis sets, and simply minimized the energy with respect to $\alpha $ using the resultant basis set of about $1200$ elements for all values of $a_{s}$.

\section{Results}
\subsection{Spectrum}
\begin{figure}[tbp]
\subfigure[]  {\includegraphics[width=1\columnwidth]
        {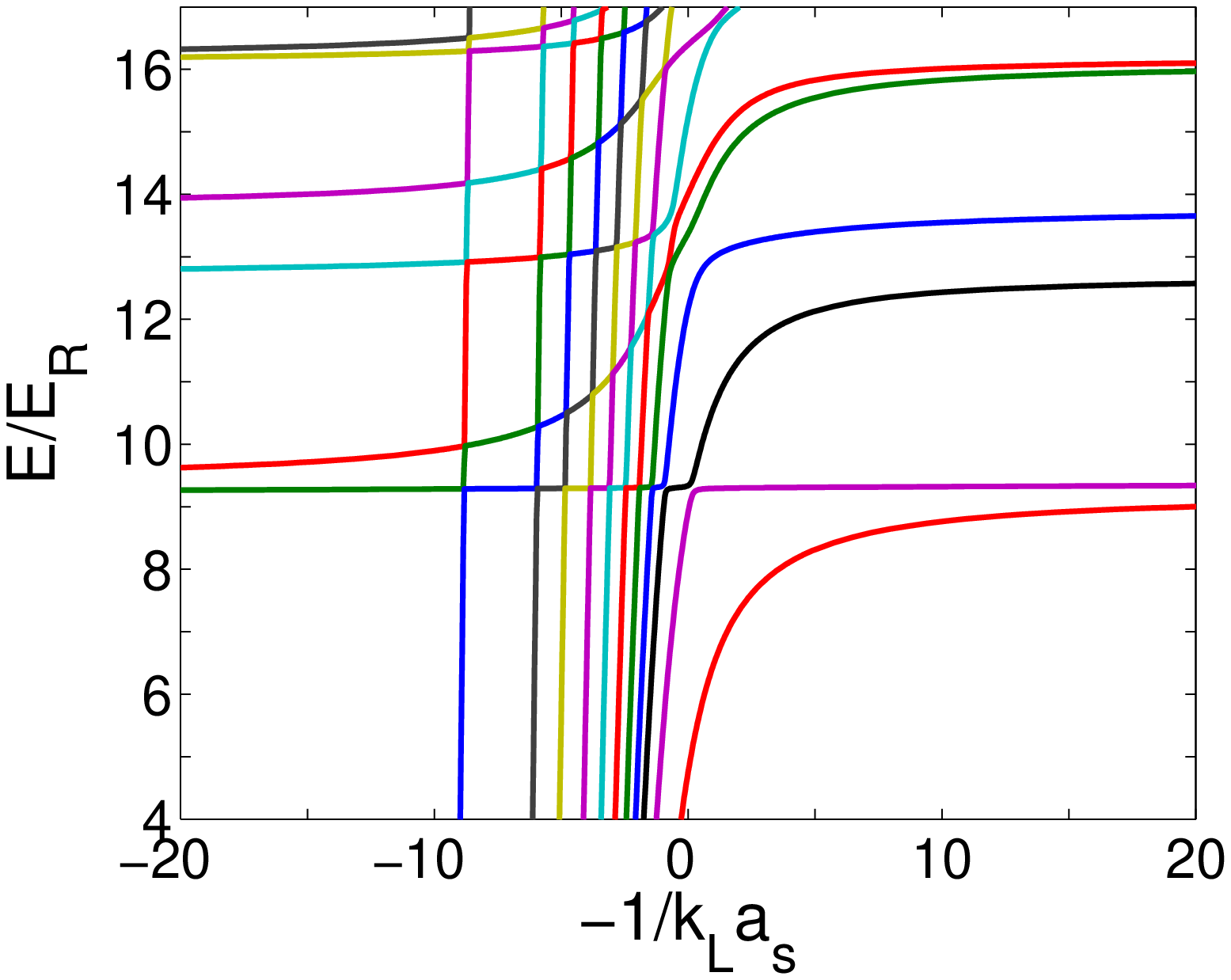}\label{fig:spectruma}}
\subfigure[]  {\includegraphics[width=1\columnwidth]
        {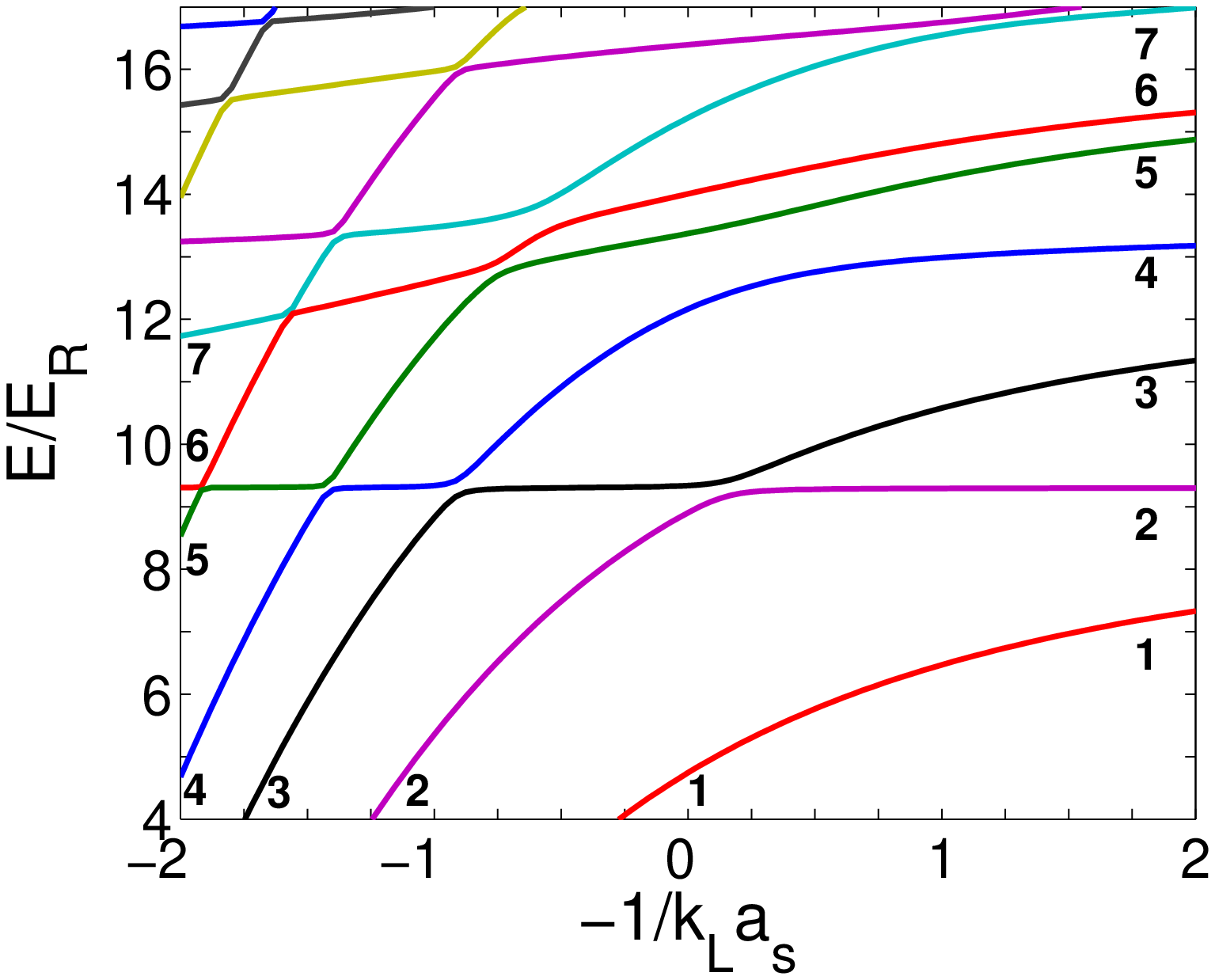}\label{fig:spectrumb}}
\caption{(Color online.) (a) Spectrum of two strongly interacting
atoms in a three-dimensional double-well potential with
$V_{0}=6E_{R}$. Only the first few plunging levels are shown. (b)
Close-up of the strongly interacting region.  The first few states are explicitly labeled for reference.} \label{fig:spectrum}
\end{figure}
In Fig.~\ref{fig:spectrum}, we show the energy spectrum of two particles
interacting near a free space Feshbach resonance ($1/k_{L}a_{s}=0$) in the
double-well potential.
For clarity, we have omitted the levels corresponding to wavefunctions of
odd parity in $z$ or $Z$ (which have no contribution to the anharmonicity induced resonances) and plunging levels for $-1/k_{L}a_{s} <-10$.  We have explicitly labeled the lowest few states for later reference.  To understand Fig.~\ref{fig:spectrum}, it is useful to use the language of the two-channel picture of atom pairs coupled to molecules, as in Fig.~\ref{fig:idea}.  Without coupling, there are plunging molecular levels crossing flat (i.e., noninteracting) atom pair levels, as depicted by the dashed lines in Fig.~\ref{fig:ideatrap}.  When one turns on atom-molecule coupling only between molecules and atoms with the same c.m.\ motion, the crossings between the lowest plunging molecular level and the flat atomic levels become avoided crossings and the spectrum is similar to the well-known results for a harmonic trap \cite{Busch98}.  The atoms and the lowest c.m.\ molecules hybridize, such that as the inverse scattering length is adiabatically swept from negative values to positive values, the lowest atomic level evolves into the lowest molecular level and a given excited atomic level will evolve into the next lowest atomic level, sweeping out a sigmoidal path.  (For a double-well, the lowest two atomic levels form a closely spaced doublet, so in the presence of coupling the lowest sigmoidal level is essentially flat, as in Fig.~\ref{fig:spectruma}.  Higher lying doublets behave likewise.)

If we take anharmonicity into account by allowing coupling also between the atoms and excited c.m.\ molecules, the crossings between the higher plunging molecular levels and the flat atomic levels also become narrow avoided crossings. These signal the presence of a rich set of induced resonances. The resonances are weak relative to the free space resonance, and become progressively weaker away from $1/k_{L}a_{s}=0$, so that only the first few are observable.  Diabatically, then, Fig.~\ref{fig:spectrum} displays three kinds of curves: plunging levels corresponding to tightly
bound molecules, flat levels corresponding to atoms in separate wells, and
sigmoidal levels corresponding to atoms with overlapping wavefunctions such that they interact while maintaining a nonvanishing pair size unlike the tightly bound molecules.  Note that the many plunging molecular levels, of which we have shown only the first few, are associated with the various states of the trap, as sketched in Fig.~\ref{fig:ideatrap}.  They are \emph{motionally} excited c.m.\ states, not internally excited states of the interaction potential.  Also note that for an optical lattice, one will obtain a similar spectrum except that the discrete levels of the double-well shown in Fig.~\ref{fig:spectrum} (analogous to Fig.~\ref{fig:ideatrap}) will broaden into bands (analogous to Fig.~\ref{fig:idealattice}).

\subsection{Avoided crossing data}
To characterize the anharmonicity induced resonance, we estimate the time
required to adiabatically sweep across the avoided level crossing,
transferring population between atomic and molecular states. In the
Landau-Zener approximation \cite{Zener}, the probability of an adiabatic
transfer at sweep rate $\partial B/\partial t = v$ is $P_{ad}=1-\exp \left( -v_{LZ}/v\right) $,
where the Landau-Zener parameter $v_{LZ}=\pi \Delta ^{2}/2\hbar |\partial
{\Delta }/\partial {B}|$, $\Delta $ is the minimum energy gap between the two
levels in question, and $\partial {\Delta }/\partial {B}$ is the rate at
which the energy gap changes with the magnetic field away from the avoided
crossing. The energy splitting $\Delta $ for the avoided level crossing
should be proportional to the width of the corresponding anharmonicity induced resonance in a periodic optical lattice. This parameter $\Delta $ is
listed in Table \ref{table:table1} for the various avoided crossings between
excited molecular states and the lowest atomic level (near $E=9.3 E_R$) for $^{6}$Li or $^{40}$K
atoms.  The numbers quoted are of course only a rough guide to what may be expected in experiment, and are not intended to be quantitatively precise -- recall that the double-well potential we have taken is only an approximation to whatever form the actual potential may take.  To connect our results to experiment, we assume the scattering length
is related to the magnetic field via the usual relation $a_{s}=a_{bg}\left[1-
W/\left( B-B_{0}\right) \right] $, where $a_{bg}$ is the background
scattering length, $W$ is the resonance width, and $B_{0}$ is the resonance
point. We take $k_{L}\sim 2\pi /1\mu m$ and consider $^{6}$Li ($^{40}$K)
near the free space Feshbach resonance at $834$ G \cite{Bartenstein05} ($202$
G \cite{Jin04}). In Table \ref{table:table1}, we have also listed an
estimate of the minimum time, $t_{min}$, required to ramp across the avoided
crossing at the critical rate, $v_{LZ}$. If the time available in the
experiment to perform the ramp is on the order of a few milliseconds,
appreciable adiabatic transfer is feasible across the first five (four)
avoided crossings for $^{6}$Li ($^{40}$K) atoms.
\begin{table}
  \begin{center}
  \begin{tabular}{|c|c|c|c|c|c|}
    \hline
     & $-1/k_L a_s$ & $\Delta /h$ (kHz) & $t_{min}$ ($\mu$s) & $v_{LZ}$ (G/s) \\
    \hline
    \multirow{5}{*}{$^{6}$Li}
    & 0.2  & 8 & 40                & $2 \times 10^6$ \\
    & -0.9 & 8 & 70                & $3 \times 10^5$ \\
    & -1.4 & 5 & 100               & $6 \times 10^4$ \\
    & -1.9 & 1  & 200               & $2 \times 10^3$ \\
    & -2.5 & 0.3  & $1 \times 10^{3}$ & 40 \\
    \hline
    \multirow{5}{*}{$^{40}$K}
    & 0.2  & 1      & 600               & 800 \\
    & -0.9 & 1      & 600               & 400 \\
    & -1.4 & 0.8      & $1 \times 10^{3}$ & 100 \\
    & -1.9 & 0.2      & $2 \times 10^{3}$ & 6 \\
    & -2.5 & 0.05    & $1 \times 10^4$   & 0.1 \\
    \hline
  \end{tabular}
  \end{center}
    \caption{Anharmonicity induced avoided level crossing data for
    $^{6}$Li ($^{40}$K) atoms at $V_{0}=6E_{R}$.  These are the avoided crossings near $E = 9.3 E_R$ shown in Fig.~\ref{fig:spectrum}.} \label{table:table1}
\end{table}

\begin{figure}[tbp]
\subfigure[]  {\includegraphics[width=.49\columnwidth]
        {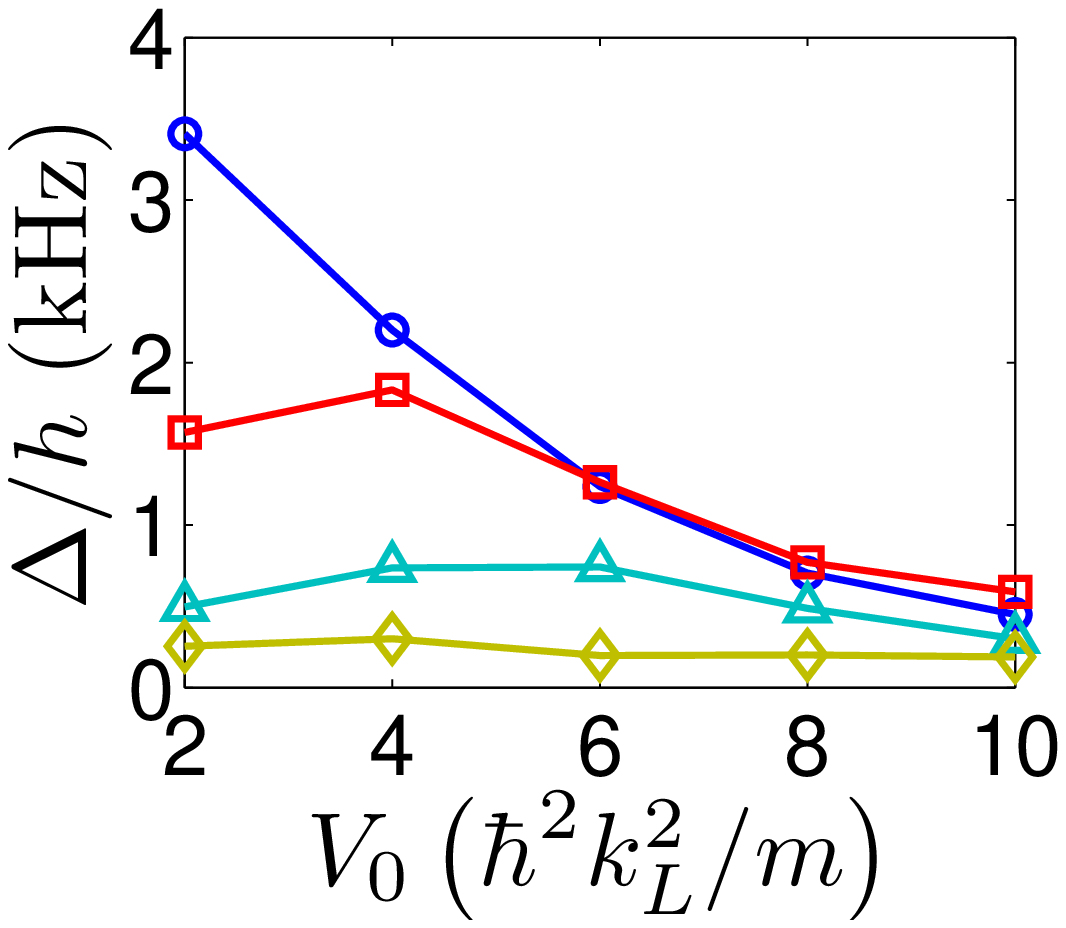}\label{fig:EgapvsV}}
\subfigure[]  {\includegraphics[width=.49\columnwidth]
        {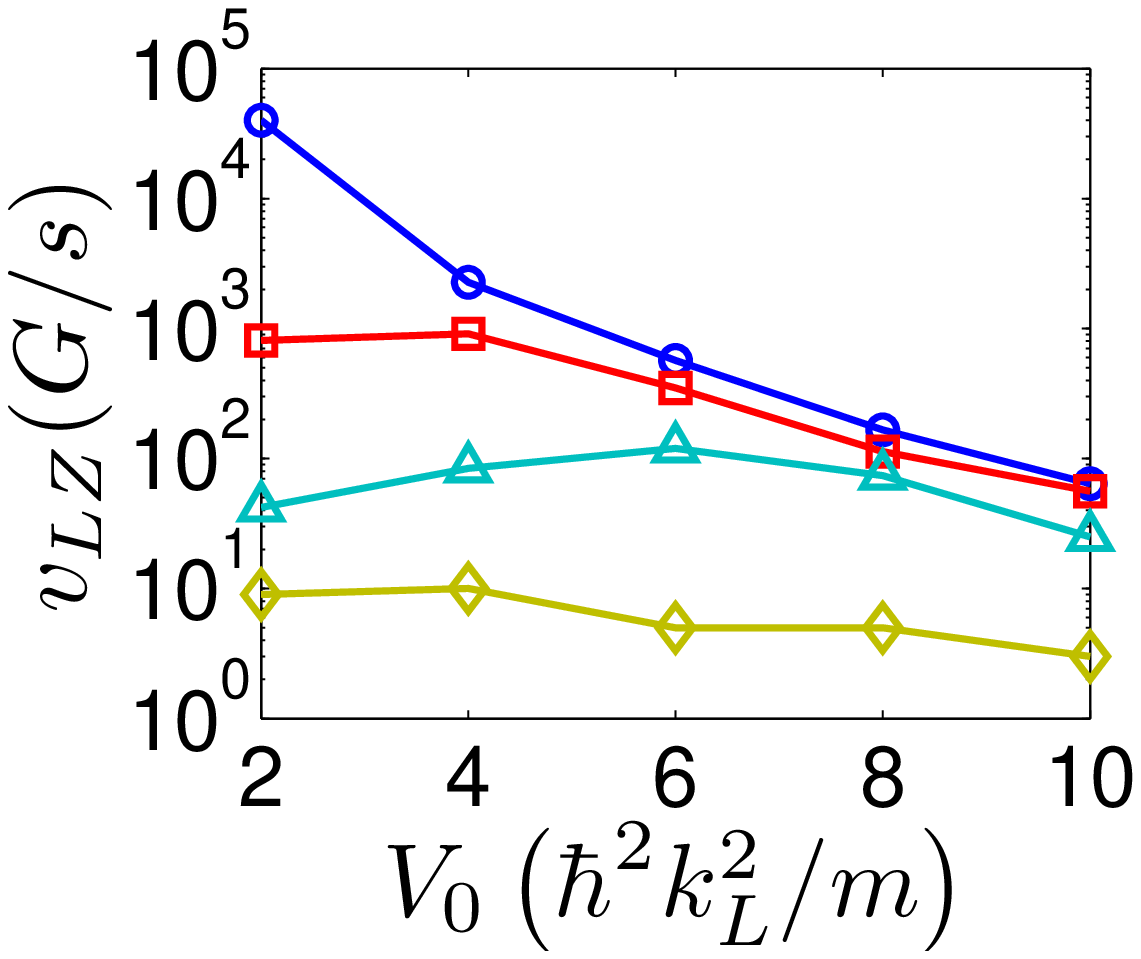}\label{fig:vLZvsV}}
\caption{(a) Energy gap and (b) Landau-Zener parameter for the
first four avoided crossings vs. well depth for $^{40}$K atoms.
From the top to the bottom, the curves correspond to the avoided crossings shown in Fig.~\ref{fig:spectrumb} at $E = 9.3 E_R$ and $-1/k_L a_s = 0.2, -0.9, -1.4, -1.9$, respectively.} \label{fig:vsV}
\end{figure}
We have performed the same kind of calculations for several lattice depths.
In Fig.~\ref{fig:vsV} we show how the energy splitting $\Delta $ and the
Landau-Zener parameter $v_{LZ}$ for the first few resonances listed in Table
\ref{table:table1} change as $V_{0}$ is varied for $^{40}$K. Generally, the
energy splitting for the avoided level crossing decreases for deeper wells,
as one would expect due to suppression of the anharmonicity in a deep
lattice (the harmonic approximation becomes better for a deep lattice well).
For very shallow wells, though, the potential
apparently cannot couple the higher c.m.\ states of Feshbach molecules to
the lowest atomic state as efficiently, and the energy splitting $\Delta $
actually increases with the lattice depth at first for small $V_{0}$. However, for very excited c.m.\ states (corresponding, e.g.\, to the bottom curve in Fig.~\ref{fig:vsV}), the weak coupling to the lowest atomic state evidently does not depend as strongly on the lattice depth. As the
potential wells are deepened, the resonance positions shift slightly to lower magnetic fields.

\subsection{Detection}
To experimentally detect the avoided level crossings associated with the
anharmonicity induced resonances, one can take the following steps: First,
one loads the optical superlattice in the weakly interacting region with two
atoms in each double well \cite{dblwell, dwell}.
The inter-well barrier is kept high so that one has a Mott state with one
atom per well. Second, one ramps the system to the strongly interacting
region with $-1/k_{L}a_{s}=\pm 2$, and then quickly lowers the
inter-well barrier to the desired value (with $V_{0}=6E_{R}$ in our
example), leaving the atoms still in the Mott state (at energy $E\simeq 9.3 E_R$ in Fig.~\ref{fig:spectrum}) at this moment. The magnetic field is then adiabatically ramped across the anharmonicity induced
resonances, and one detects the resulting population distribution after the
ramp. To do the detection, the inter-well barrier is quickly turned back up with a time scale fast compared with the inter-well dynamics, but still slow compared with the lattice band gap (or the single-well energy gap).
This freezes the system evolution again before the magnetic field is ramped to
the deep BEC side ($-1/k_{L}a_{s}\ll -1$), separating the molecular levels
from the atomic levels. One can then selectively take absorption images of
either the atoms or the molecules \cite{Jin03}, and measure their
distribution over different bands through a band-mapping procedure
\cite{dblwell}. The presence of the anharmonicity induced level crossings
can then be inferred from the final population distribution.

\begin{figure}[tbp]
\subfigure[]  {\includegraphics[width=.49\columnwidth]
        {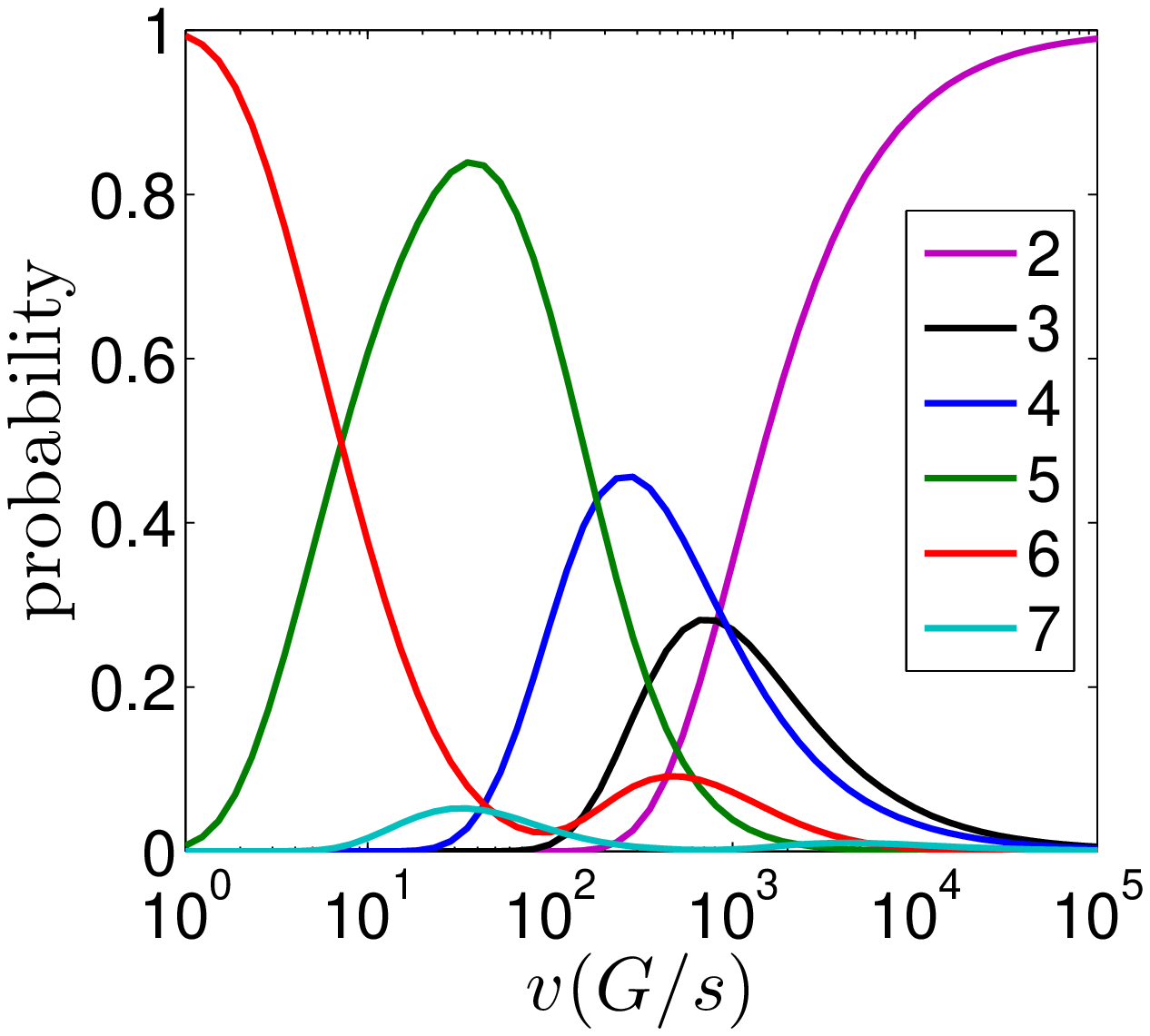}\label{fig:sweepa}}
\subfigure[]  {\includegraphics[width=.49\columnwidth]
        {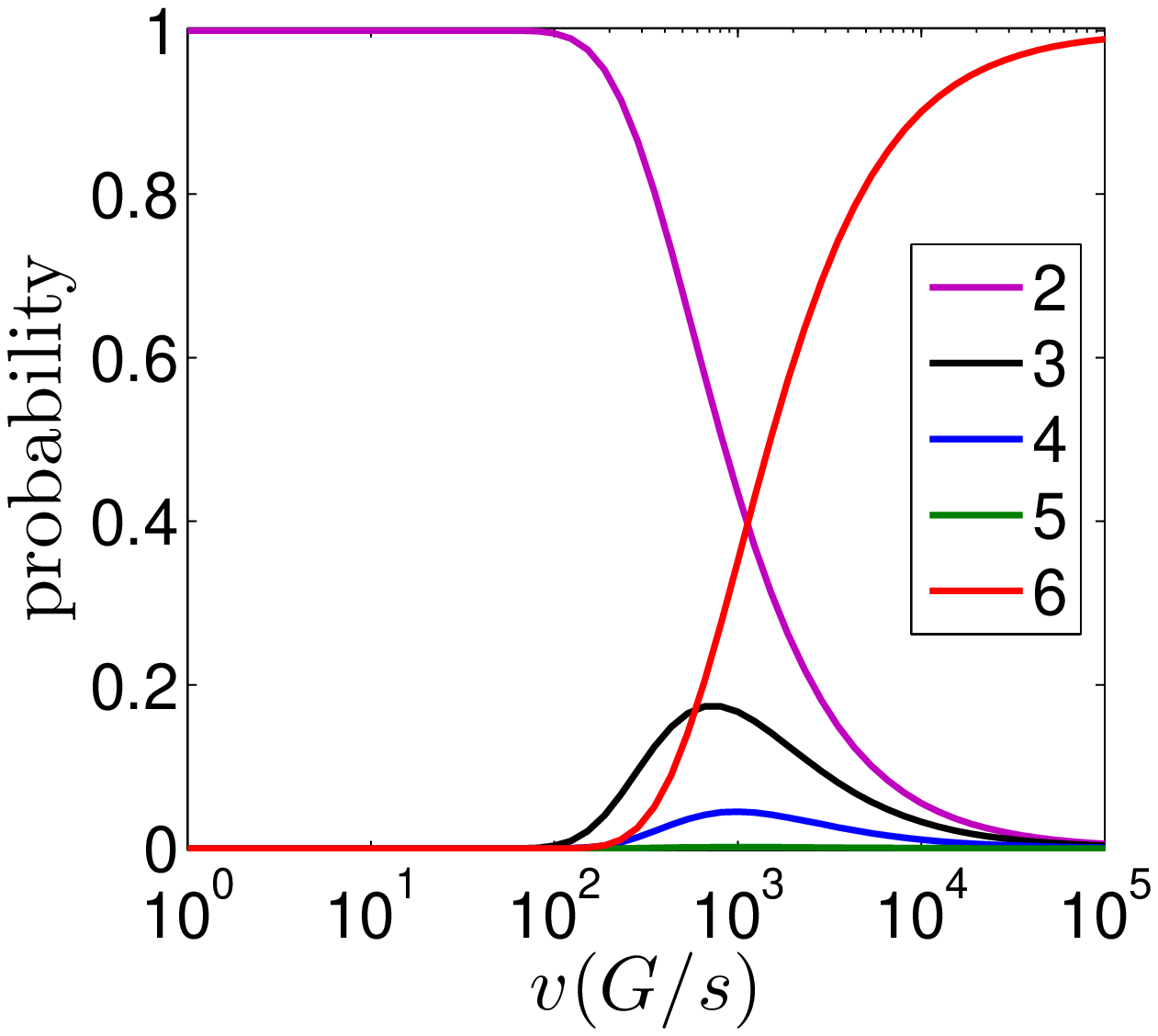}\label{fig:sweepb}}
\caption{(Color online.) Final population distribution vs. ramp
speed of the magnetic field (a) from state 6
at $-1/k_{L}a_{s}=-2$ to states 2-7 at $-1/k_{L}a_{s}=2$ or (b) from state 2 at
$-1/k_{L}a_{s}=2$ to states 2-6 at $-1/k_{L}a_{s}=-2$. Both plots are for $^{40}$K atoms with
$V_{0}=6E_{R}$.  In (b), the probability of sweeping into state 5 at $-1/k_{L}a_{s}=-2$ is essentially zero at any ramp speed.} \label{fig:sweep}
\end{figure}
As an example, in Fig.~\ref{fig:sweepa}, we show the Landau-Zener calculation results for $^{40}$K atoms at $V_{0}=6E_{R}$ swept from
$-1/k_{L}a_{s}=-2$ to $-1/k_{L}a_{s}=2$ with the atoms starting in the state 6 labeled in Fig.~\ref{fig:spectrumb}. For fast sweeps, the atoms change states diabatically to remain at about the same energy, as would be expected in the absence of anharmonicity. In the adiabatic limit, all the atoms remain in state 6, which corresponds at $-1/k_{L}a_{s}=2$ to atoms in an excited state. At
intermediate speeds, several atomic states become populated. A sweep
in the opposite direction, from $-1/k_{L}a_{s}=2$ to $-1/k_{L}a_{s}=-2$,
starting with atoms in state 2, is shown in
Fig.~\ref{fig:sweepb}. When sweeping in this direction, population can be transferred
to tightly bound molecules in several excited c.m.\ states (states 2-5 at $-1/k_{L}a_{s}=-2$) as well as diabatically to atoms near the initial energy (state 6).

\section{Summary}
We predict the existence of several Feshbach-type resonances
induced by the anharmonicity of the optical lattice, which couples the Feshbach molecules in the excited
bands and the atomic states in the lowest band. We have characterized the
corresponding set of avoided level crossings in the calculated spectrum of
two atoms interacting in a superlattice potential, and proposed an
experimental scheme to observe these avoided crossings through slow sweeps
of the magnetic field. The anharmonicity induced resonances may prove to be
a useful tool for manipulation of interaction between ultracold atoms in
optical lattice potentials.

This work was supported by the AFOSR\ through MURI, the DARPA, and
the IARPA.

\end{document}